\documentclass{aastex}
\usepackage{emulateapj5}

\newcommand{\ie}{\hbox{i.e.}}
\newcommand{\etal}{\hbox{et~al.}}
\newcommand{\Mdot}{\hbox{$\dot M$}}
\newcommand{\Msunyr}{\hbox{$M_\odot\,$yr$^{-1}$}}
\newcommand{\Rsun}{\hbox{$R_\odot $}}
\newcommand{\vinf}{\hbox{$v_\infty$}}
\newcommand{\zori}{\hbox{$\zeta$ Ori}}

\newcommand{\zpup}{\hbox{$\zeta$ Pup}}
\newcommand{\kms}{\hbox{km$\,$s$^{-1}$}}

\newcommand{\Rstar}{\hbox{R$_*$}}

\newcommand{\Lyalpha}{\hbox{Ly$\alpha$}}

\newcommand{\Teff}{\hbox{$T_{\rm eff}$}}

\providecommand{\OVII}{\ion{O}{7}}
\providecommand{\OVIII}{\ion{O}{8}}
\providecommand{\NEIX}{\ion{Ne}{9}}
\providecommand{\NEX}{\ion{Ne}{10}}
\providecommand{\MGXI}{\ion{Mg}{11}}
\providecommand{\MGXII}{\ion{Mg}{12}}
\providecommand{\SIXIII}{\ion{Si}{13}}

\providecommand{\SXV}{\ion{S}{15}}

\providecommand{\NVII}{\ion{N}{7}}
\providecommand{\FEXVII}{\ion{Fe}{17}}

\newcommand{\fir}{\hbox{\it fir}}

\newcommand{\einstein}{\hbox{\it Einstein}}
\newcommand{\chandra}{\hbox{\it Chandra}}

\newcommand{\xmm}{\hbox{\it XMM}}

\newcommand{\bbxrt}{\hbox{\it BBXRT}}
\newcommand{\smallf}{\hbox{f}}

\shorttitle{X-ray Source Properties of O-star \zpup}
\shortauthors{Cassinelli \etal}

\begin{document}

\title{{\it Chandra} Detection of Doppler Shifted X-ray Line Profiles
from the Wind of $\zeta $ Puppis~ (O4\smallf)}
				   
\author{J. P. Cassinelli\altaffilmark{1} ,
N. A. Miller\altaffilmark{1}, W. L. Waldron\altaffilmark{2},
J. J. MacFarlane\altaffilmark{3}, and D.  H. Cohen\altaffilmark{3,4}}
\altaffiltext{1}{Astronomy Department, University of Wisconsin, 475
N. Charter St., Madison, WI  53706; cassinelli@astro.wisc.edu,
nmiller@astro.wisc.edu} \altaffiltext{2}{Emergent Information
Technologies, Inc., 9314 Largo Drive West, Suite 250, Largo, MD 20774;
wayne.waldron@emergent-IT.com} \altaffiltext{3}{Prism Computational
Sciences, 16 N. Carroll St. Madison WI 53703; jjm@prism-cs.com,
cohen@prism-cs.com} \altaffiltext{4}{Department of Physics and
Astronomy, Swarthmore College, Swarthmore, PA 19081;
dcohen1@swarthmore.edu}

\begin{abstract}

We report on a 67 ks HETG observation of the optically brightest early
O-star, \zpup\ (O4 f). Many resolved X-ray lines are seen in the
spectra over a wavelength range of 5 to 25 \AA.  \chandra\ has
sufficient spectral resolution to study the velocity structure of
isolated X-ray line profiles, and to distinguish the individual
forbidden, intercombination, and resonance ($fir$) emission lines in
several He-like ions even where the individual components are strongly
Doppler broadened.  In contrast with X-ray line profiles in other hot
stars, \zpup\ shows blue-shifted and skewed line profiles, providing
the clearest and most direct evidence that the X-ray sources are
embedded in the stellar wind. The broader the line, the greater the
blueward centroid shift tends to be.  The \NVII\ line at 24.78 \AA\ is
a special case, showing a flat-topped profile.  This indicates it is
formed in regions beyond most of the wind attenuation.  The
sensitivity of the He-like ion $fir$ lines to a strong UV radiation
field is used to derive the radial distances at which lines of \SXV,
\SIXIII, \MGXI, \NEIX, and \OVII\ originate. The formation radii
correspond well with continuum optical depth unity at the wavelength
of each line complex, indicating that the X-ray line emission is
distributed throughout the stellar wind.  However, the \SXV\ emission
lines form deeper in the wind than expected from standard wind shock
models.

\end{abstract}

\keywords{ X-rays: stars ---  stars: individual (\zpup) --- stars:
early-type --- stars: winds, outflows --- stars: mass-loss --- line:
profiles }

\section{Introduction}

The O4f star $\zeta$ Puppis has for decades been at the cutting edge
of research regarding early type stars because of its optical and UV
brightness.  In addition, because of the low interstellar attenuation,
\zpup\ has been the prime X-ray target to study soft X-ray emission
from O stars.  Researchers have looked to this source to settle
outstanding controversies about the physical location, quantity, and
nature of the hot X-ray emitting plasma on OB stars.  In this paper we
present the highest resolution X-ray spectrum of \zpup\ ever measured
and explore  the physical mechanism of hot star X-ray production.

Prior to the discovery of X-ray emission from O stars (Harnden et
al. 1979; Seward et al. 1979), Cassinelli \& Olson (1979) postulated
thin base coronal zones as a source for X-rays to explain the observed
UV superionization (Lamers \& Morton 1976) by way of the Auger effect.
The coronal models (Cassinelli \& Olson 1979; Waldron 1984) predicted
an X-ray absorption at the oxygen K-shell edge larger than that
observed with \einstein\ SSS (Cassinelli \& Swank 1983) and \bbxrt\,
(Corcoran \etal\ 1993). These observations, as well as detailed
modeling of the superionization profiles (MacFarlane \etal\ 1993) and
lack of detection of the iron ``green line'' (Baade \& Lucy 1987) in
\zpup\, caused the coronal model to fall out of favor.  Increasingly,
a consensus formed around a wind-shock picture in which a series of
shocks, perhaps related to the line-force instability (Lucy 1982;
Owocki, Castor, \& Rybicki 1988), causes hot, X-ray emitting gas to be
distributed {\it throughout} the dense stellar wind of \zpup\ and
other OB stars.  Wind shock models developed by Lucy \& White (1980),
Feldmeier \etal\ (1997) and others, consistently failed to predict the
high levels of X-ray emission observed in the brightest O stars like
\zpup, leading to the suggestion that perturbations somehow form and
propagate up from the photosphere into the wind and drive stronger
shocks (Feldmeier 1995; Feldmeier, Puls, \& Pauldrach
1997). Broad-band X-ray observations of \zpup\/ (Corcoran et al. 1993;
Hillier et al. 1993) indicate that some wind attenuation is affecting
the soft X-ray flux. However with the advent of \chandra\/ and \xmm\/
we can apply, for the first time, diagnostic emission line ratios and
measure line profiles to determine the locations and Doppler
velocities of X-ray sources in the stellar wind of \zpup.

\section{Observations and Constraints from Helium-like {\it fir} Line Ratios}

We obtained a 67 ks \chandra\ High Energy Transmission Grating
Spectrometer (HETGS) observation of \zpup\ from 2000 Mar 28, 13$^{h}$
31$^{m}$ UT to Mar 29, 09$^{h}$ 12$^{m}$.  The standard pipeline tools
were used to reprocess the data with the most recent calibration files
available. Line emission is clearly evident in both the high-energy
grating (HEG) and the medium-energy grating (MEG).  The combined $\pm
1^{st}$ order, background-subtracted, HETGS spectra are shown in
Figure 1.  Triads of He-like ions known as the \fir\ (forbidden,
intercombination, and resonance) lines (\SXV, \SIXIII, \MGXI, \NEIX,
and \OVII) are seen, in addition to isolated \Lyalpha\ emission lines
and numerous L-shell lines of iron, especially \FEXVII.

\resizebox{8.4cm}{!}{ \rotatebox{180}{
\includegraphics{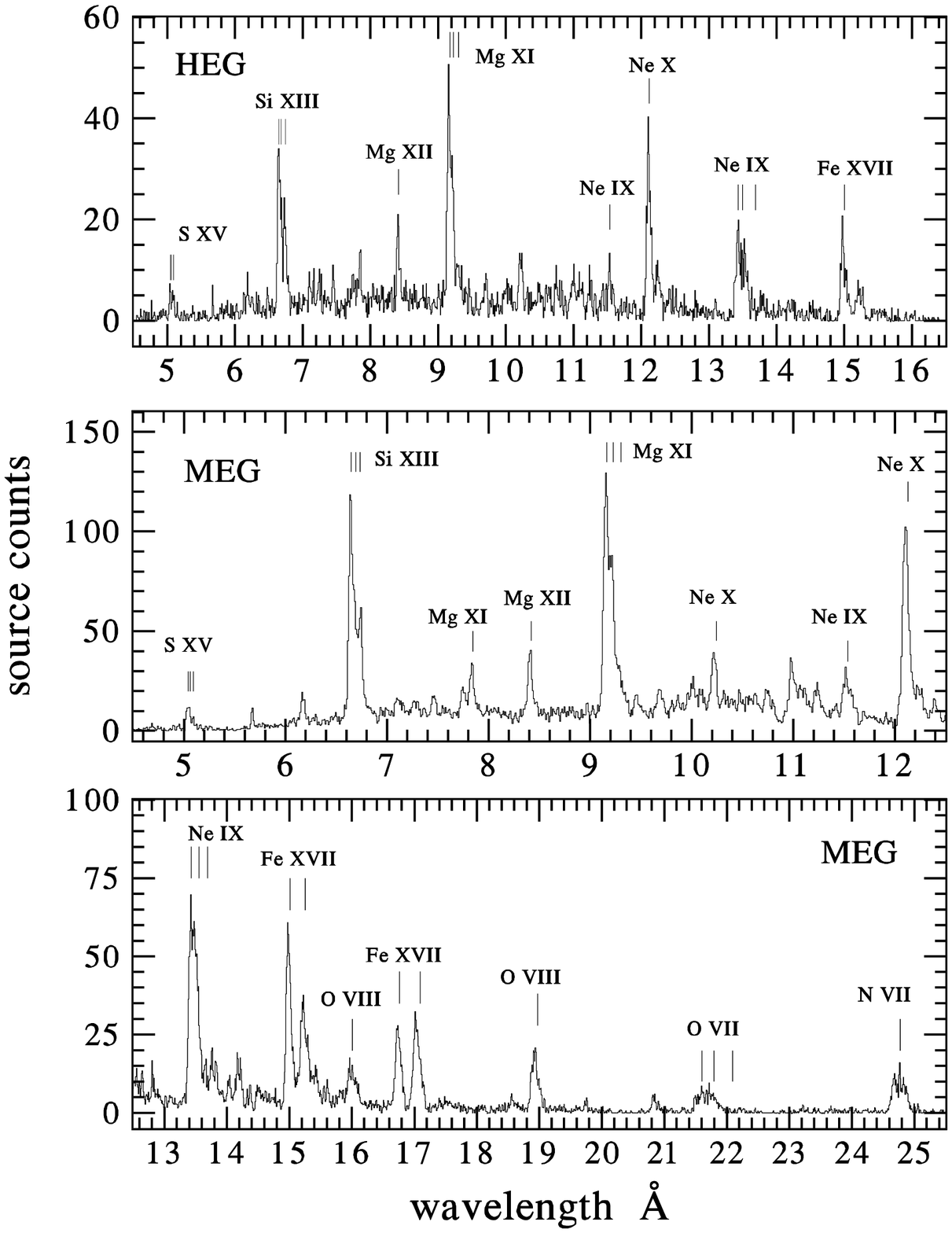}}}
\figcaption[ChandraZPupfig1.eps]{Co-added \protect$1^{st}$ order,
background subtracted, Chandra HETGS HEG (top panel) and MEG (bottom
panels) count spectra of \protect\zpup.  The ions responsible for the
strongest line emissions are identified.  The bin size is 0.01
\protect\AA. }
\vspace{0.5cm}

Traditionally, He-like ion $f/i$ line ratios have been used
to derive electron densities of X-ray line emitting
regions since the populations of the 2$^3$P levels are controlled by
collisional excitations from the 2$^3$S levels (Gabriel \& Jordan
1969).  However, when a strong external UV source is present, the
excitation 2$^3$S $\rightarrow$ 2$^3$P  is predominantly radiative and
this means that the $f/i$ ratio is no longer a density diagnostic, but
rather it can be used to determine the strength of the UV radiation
field (Blumenthal \etal\ 1972).  The UV wavelengths associated with
the radiative excitation are in the range from $\sim\ 650 $ to 1650
\AA, where O star UV fluxes are very strong.  For \zpup's photospheric
flux we use a Kurucz (1993) model, assuming \Teff = 42000K.  Since the
 2$^3$S $\rightarrow$ 2$^3$P wavelengths are not on the Wien side
of the spectrum, the model atmosphere fluxes are relatively well determined.  

Waldron \& Cassinelli (2001) demonstrated how the observed $f/i$ line
ratios can be used to derive the line formation radii from the
geometric dilution of the photospheric radiation field in \zori\
(09.5Ia).  Kahn \etal\ (2001) used a similar approach to derive line
formation radii from their \xmm\ {\it RGS} spectrum of \zpup.  Our
\chandra\ HETG spectrum has better resolution at essentially all
He-like ion wavelengths and greater sensitivity to the
short-wavelength lines (e.g., Si and S) than the \xmm\ spectrum.
This high resolution is very important for O stars since their X-ray
lines show strong broadening.  The observed He-like $f/i$ line ratios
are listed in Table 1.

\begin{table*}[ht!]
{\bf Table 1:  $fir$ ratio measurements }
\vspace{-0.1cm}
\begin{center}
\begin{tabular}{llllll}
\tableline
\tableline
       & \protect\SXV & \protect\SIXIII & \protect\MGXI & \protect\NEIX   & \protect\OVII \\
\tableline
{ \it f/i }  &$  0.61  \pm 0.48  $&$ 1.03 \pm 0.14  $&$ 0.29 \pm 0.18
$& $<$ 0.1 & $<$ 0.1 \\
{ \it r/i }  &$     0.78    \pm   0.54   $&$    2.00
\pm   0.25  $
&$  1.53   \pm  0.13 $&$  1.44 \pm  0.18 $&$1.30  \pm  $   0.6 \\
${\cal F}_{tot} \tablenotemark{a}$& 0.81 &3.06&3.26&5.77&3.56\\
\tableline
\end{tabular}
\end{center}
\footnotesize
\vspace{-0.3cm}
\hspace{3.7cm}
\protect$^{a}$Total flux of the three lines in \protect$10^{-13}$ erg cm\protect$^{-2}$ s\protect$^{-1}$
\end{table*}

Figure 2a shows the radii derived from our measurements of the He-like
$f/i$ ratios. The radii of line formation are seen to vary from ion to
ion, with lines from high nuclear charge ions (Si, S) forming at small
radii, and lower nuclear charge ion lines (O, Ne, Mg) forming at
larger radii.  One might expect the $fir$ line complexes for each
element to be formed over a range of radii and not just in a thin
shell. However, since the line emission is proportional to $n_{\rm
e}^2$, the observed emission in each line complex should be dominated
by the densest regions (\ie\ the smallest radius) from which the line
radiation can escape.  For \zori\ the $fir$ formation locations
correlates with the associated radial optical depth unity radius,
$R_1$. The optical depth is measured through the wind, using the
continuum opacity at the wavelength of each triad of $fir$ lines.  Any
line radiation that originates much below $R_1$ could not be seen by
an external observer.  The wavelength dependence of $R_1$ for \zpup\
is shown in Figure 2b using a $\beta$-law with $\beta=$ 0.75 as
the velocity law, the wind absorption cross sections in Waldron \etal\
(1998), and the stellar parameters of Lamers \& Leitherer (1993)
($R_*$ = 16 \Rsun, \Mdot= $2.4 \times 10^{-6}$ \Msunyr, and \vinf =
2200 \kms).

\resizebox{8.4cm}{!}{ \rotatebox{180}{
\includegraphics{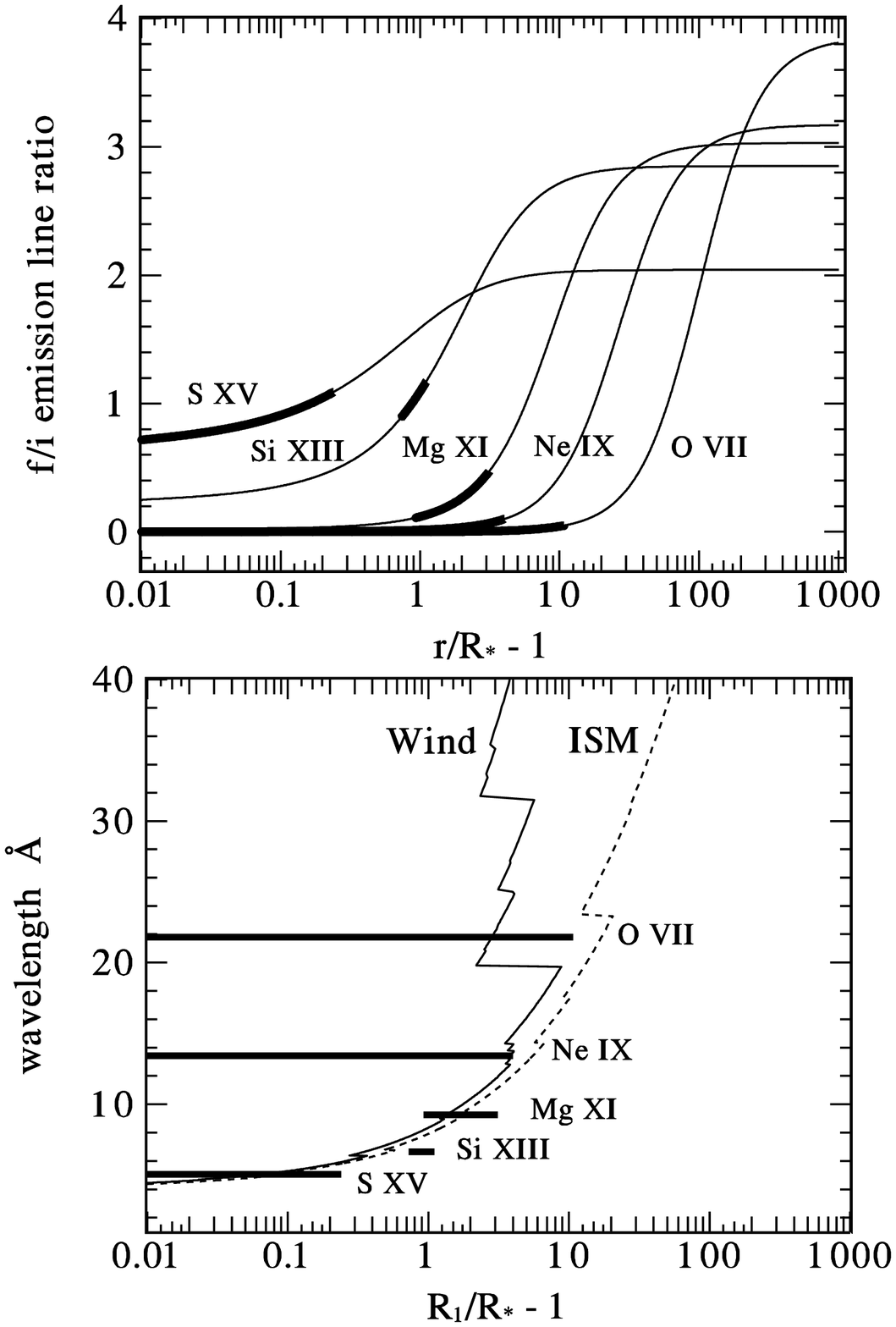}}}
\figcaption[ChandraZPupfig2.eps]{The top panel shows the dependence of
\protect\OVII, \protect\NEIX, \protect\MGXI, \protect\SIXIII\ and
\protect\SXV\ \protect$f/i$ ratios on radius due to the geometric
dilution of the UV radiation field. The length of the darkened line
for each ion extends over the range of the uncertainty in the \protect$f/i$
line ratio. The bottom panel gives the radius of wind continuum
optical depth unity as a function of wavelength.  The dark horizontal
lines correspond to the range in radii indicated in the upper panel.
For comparison, we also show the optical depth unity radius assuming
neutral interstellar medium opacity of Morrison \& McCammon (1983). }
\vspace{0.5cm}

An inspection of Figure 2 leads to the following conclusions: First,
the He-like line formation radii are all consistent with their
associated $R_1$ values, as expected from the above argument.  Second,
the $f/i$ ratio of \SXV\ indicates that hot plasma exists below 1.2
\Rstar\ where the nominal wind velocity is $<$ 600 \kms.  Most
radiative instability simulations do not show strong shocks this close
to the star (Feldmeier 1995, Owocki, Castor, \& Rybicki 1988).
However, using a velocity jump even half as large as the local wind
velocity ($\Delta v = 300 \kms$) to calculate the shock temperature
from $ T = 1.4 \times 10^5 K (\Delta v/100 \kms)^2$, we obtain a
post-shock temperature of 1 MK; too low to produce \SXV. Thus,
anomalously strong shock jumps are required deep in the wind. A
similar conclusion was reached by Waldron \& Cassinelli (2001), but
for \zori\ the highest ion stage observed was \SIXIII\ instead of the
\SXV\ seen here.

\section{Emission Line Profiles}

The most notable feature observed in our HETG spectrum of \zpup\ is
the clear presence of blue-shifted X-ray line centroids in all strong
lines.  This is in contrast with previous \chandra\ O star observations
(Schulz \etal\ 2000; Waldron \& Cassinelli 2001) where the lines were
broad, symmetric and un-shifted. Although our line widths are
comparable to other O-star observations, with HWHM of $\sim$ 1000
\kms, our observation is the first \chandra\ detection of blue-shifted
and blueward skewed X-ray line profiles in an O star. Figure 3
displays six \zpup\ X-ray lines arranged in order of wavelength.  The
results for the line fits are given in Table 2.  There is an
interesting progression that holds for these lines: the centroid
shifts are generally larger for the broader lines, indicating a
connection between radial location, wind absorption, and the Doppler
broadening of the line emission. In Table 4 of their XMM study of
\zpup, Kahn \etal\ (2001) list blueshifts of the Lyman-$\alpha$ lines
of \NEX\ and \OVIII. The blueshifts agree with ours to within 2 sigma,
and some of the discrepancy can be attributed to the absolute
wavelength calibration uncertainties of the instruments ($\sim$ 100 km
s$^{-1}$ for the HETG (Chandra X-ray Center (CXC) 2001), see den
Herder \etal\ 2001 for the RGS).  The XMM results show the same trend
between shift and broadening.  The \NVII\ line is an exception to the
trend in Table 2 because of its small centroid shift.  It is different
morphologically from the other lines (also noted by Kahn
\etal\ (2001)).  It has a roughly flat-topped profile so we use a box
fit instead of a Gaussian to estimate the values for Table 2.

\begin{table*}[ht]
{\bf Table 2:  Line Profile Widths and Centroid Shifts}
\vspace{-0.1cm}
\begin{center}
\begin{tabular}{llllll}
\tableline
\tableline
Ion & $\lambda_{\rm rest}$ & $ {\cal F}_{\rm line} \tablenotemark{a}  $ 
          & ~~Source  &
Centroid  & $R_1$   \\
  & & &~HWHM & ~~Shift & \\
    & ~~(\protect\AA)           &   & ~~(km s$^{-1})$   &
~(km/s) & ($R_{*}$) \\
\tableline
\protect\MGXII & 8.42    &$ 0.30    $&$ 610 \pm 220 $&$ -440
\pm 130 $  & 2.8      \\
\protect\NEX & 12.13   &$ 3.33 $&$ 780 \pm 60  $&$ -570
\pm 50  $ & 4.5      \\
\protect\FEXVII & 15.01   &$ 4.74 $&$ 800 \pm 70  $&$ -500
\pm 60  $ & 6.0      \\
\protect\FEXVII & 16.78   &$ 2.24     $&$ 850 \pm 130 $&$ -730
\pm 100 $ & 7.0      \\
\protect\OVIII   & 18.97   &$ 3.05     $&$ 990 \pm 201 $&$ -670
\pm 140 $ & 4.0      \\
\protect\NVII   & 24.78   &$ 5.02   $&$ 1570 \pm 150  $&$ -120
\pm 150 $ & 4.0      \\
\tableline
\footnotesize
\protect$^{a}$Line flux in \protect$10^{-13}$ ergs cm\protect$^{-2}$  s\protect$^{-1}$ 
\normalsize
\end{tabular}
\end{center}
\end{table*}

There is a blueward skewness observed in many of the \zpup\ lines
which is not attributable to the HETGS line response function (CXC
2001).  The blue sides of these lines tend to be steeper, while the
red sides have a shallower slope. In general, the shape of these lines
indicates wind broadening combined with attenuation as would arise
from line formation in the marginally thick region near
$R_1$. MacFarlane \etal\ (1991), Ignace (2001), and Owocki \& Cohen
(2001) demonstrate that emission line profiles can provide important
information about the spatial distribution of hot plasma within a
stellar wind based both on the velocity-dependence of the intrinsic
emission and also on the continuum attenuation across the
line. Assuming isotropic emission, the red-shifted emission from the
back side of the wind is suppressed by the continuum opacity along the
line of sight, while the blue-shifted emission from the near side is
less so. This leads to skewed, almost triangular-shaped lines in an
optically thick wind.

\resizebox{8.4cm}{!}{
\includegraphics{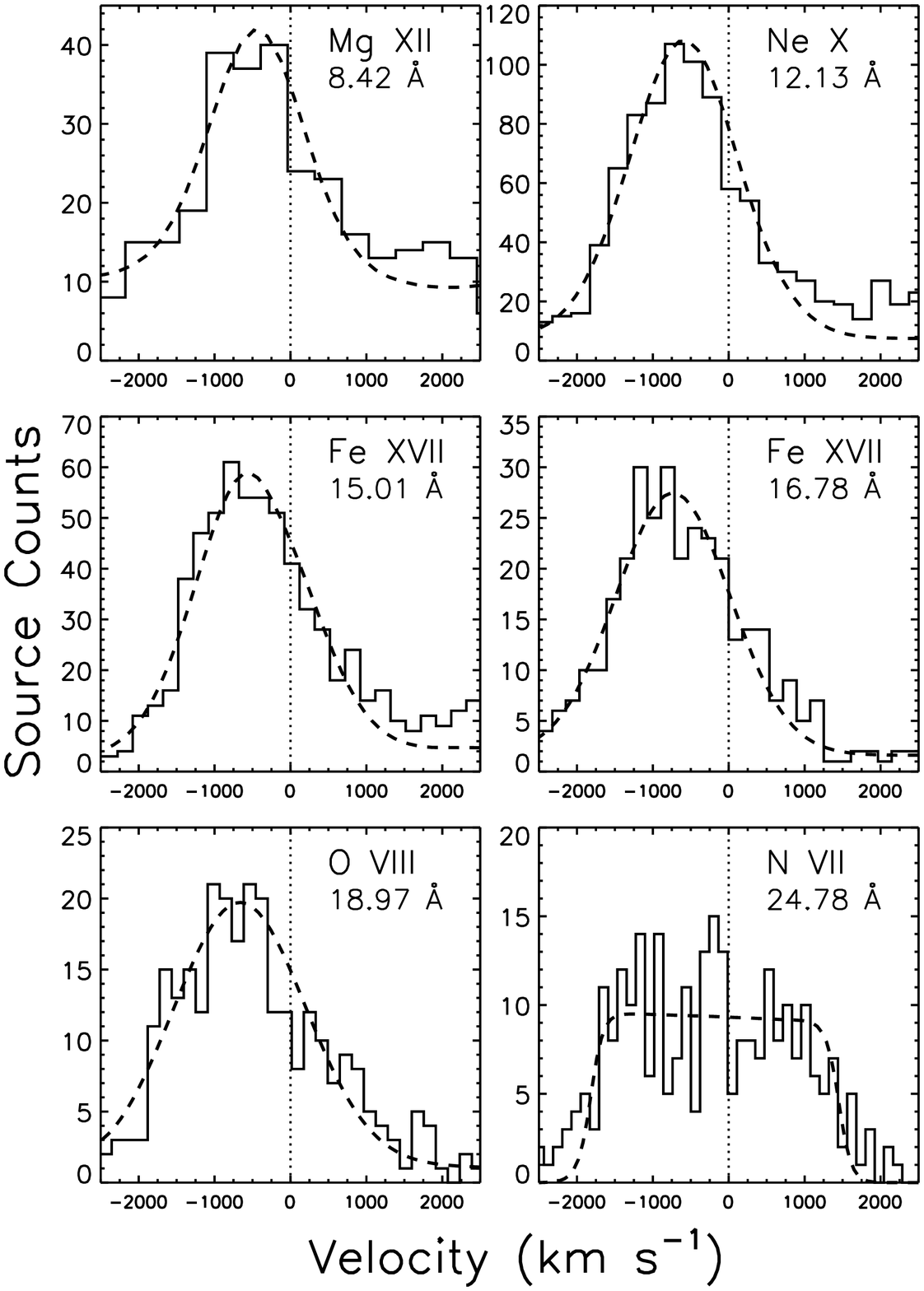}}
\figcaption[ChandraZPupfig3.eps]{Six X-ray line profiles from the
\zpup\ spectrum (solid) with fits (dashed). The vertical dotted
lines indicate the rest wavelengths for these transitions.  The first
5 lines were fit with Gaussians while the \NVII\ line was fit with a
box function because of its flat-topped shape.  All functions shown
have been convolved with the instrumental response.  These functions
are simultaneous fits to the MEG+1 and MEG-1 order spectra but are
shown here superimposed on the co-added MEG first order spectrum for
the sake of brevity.  The data have been binned to 0.01 \AA.  The fit
parameters are given in Table 2.}
\vspace{0.5cm}

In the \zori\ HETG spectrum, Waldron \& Cassinelli (2001) found that
an optically thin model was needed to provide a reasonable fit because
the lines were symmetric and un-shifted.  This fit required that the
assumed mass loss rate for \zori\ be greatly reduced.  For \zpup\ we
have the odd situation that the blueward shift and asymmetric line
profiles agree with what has long been expected for hot stars, but
this is the first time that it has been so clearly observed in a
\chandra\ spectrum for any star. We find that a fit can be made to the
line profiles without making a significant reduction in the assumed
mass loss rate in contrast with the \zori\ results.

The \NVII\ line profile is different from those of the other ions as
can be seen in Figure 3 and Table 2.  The Doppler half-width of this
line is much greater than any of the other X-ray lines.  This velocity
of 1570 \kms\ can be compared with the terminal speed of 2200 \kms\
determined from UV wind lines.  A flat-topped line profile such as
this is expected from a fast moving shell source suffering little or
no wind attenuation (MacFarlane \etal\ 1991).  Therefore, in contrast
with the other observed lines, the NVII profile is probably formed
well above its continuum optical depth unity radius.

\section{Discussion and Summary}

The \chandra\/ HETG emission line spectrum of \zpup, with its
unprecedented resolution (in excess of ${\lambda \over \Delta \lambda}
= 1000$ for some lines) and sensitivity over a wide range of
wavelengths, allows us to draw quantitative conclusions about the
nature of the X-ray source on this prototypical hot star.  Most
importantly, we find that the wind plays an important, observable role
in determining the X-ray line profiles of \zpup, in contrast to results
for other stars.

We can begin to constrain the physical processes which lead to the
production of the hot, X-ray emitting plasma on this star. Taking all
the He-like ions from oxygen through sulfur into account, the
correlation between the formation radius of the $fir$ lines and the
radius at optical depth unity, $R_1$, we find evidence for a spatially
distributed source of X-rays throughout the expanding wind.

There are two lines of particular interest in the development of
future shock models: \SXV, and \NVII. The shock jump required to
produce \SXV\ appears inconsistent with the expected local wind
conditions, requiring a post-shock flow velocity of approximately
zero. Although this is not impossible for a shock model to reproduce,
it places strong constraints on future modeling efforts.  The \NVII\
\Lyalpha\ line line may have a shape distinct from the other lines in
our spectrum because it is formed in the outer regions of the wind
where there is no further acceleration of the X-ray source region and
little overlying wind attenuation.  This \chandra\/ HETG data set
provides the most detailed and complete picture to date of an extended
distribution of hot plasma embedded within a strong, optically thick
stellar wind.  \acknowledgements

\end{document}